\documentclass[conference]{IEEEtran}
\usepackage{blindtext, graphicx}
\usepackage{textcomp}
\usepackage{colortbl}
\usepackage{subfigure}
\usepackage{algorithm}
\usepackage[noend]{algpseudocode}
\usepackage{cite}

\usepackage[cmex10]{amsmath}

\usepackage{listings}
\usepackage{color}
\usepackage{url}

\definecolor{dkgreen}{rgb}{0,0.6,0}
\definecolor{gray}{rgb}{0.5,0.5,0.5}
\definecolor{mauve}{rgb}{0.58,0,0.82}

\usepackage{etoolbox}
\makeatletter
\patchcmd{\@makecaption}
  {\scshape}
  {}
  {}
  {}
\makeatletter
\patchcmd{\@makecaption}
  {\\}
  {.\ }
  {}
  {}
\makeatother

\begin{document}
%
\title{Response of Selective Attention in Middle Temporal Area}

\author{
\IEEEauthorblockN{Linda Wang}
\IEEEauthorblockA{Systems Design Engineering\\
University of Waterloo\\
Waterloo, Canada\\
ly8wang@uwaterloo.ca}
}

\maketitle

\begin{abstract}
\boldmath
The primary visual cortex processes a large amount of visual information, however, due to its large receptive fields, when multiple stimuli fall within one receptive field, there are computational problems \cite{neural-mechanisms}. To solve this problem, the visual system uses selective attention, which allocates resources to a specific spatial location, to attend to one of the stimuli in the receptive field \cite{womelsdorf}. During this process, the center and width of the attending receptive field change. The model presented in the paper, which is extended and altered from Bobier et al. \cite{bobier}, simulates the selective attention between the primary visual cortex, V1, and middle temporal (MT) area. The responses of the MT columns, which encode the target stimulus, are compared to the results of an experiment conducted by Womelsdorf et al. on the receptive field shift and shrinkage in macaque MT area from selective attention \cite{womelsdorf}. Based on the results, the responses in the MT area are similar to the Gaussian shaped receptive fields found in the experiment. As well, the responses of the MT columns are also measured for accuracy of representing the target visual stimulus and is found to represent the stimulus with a root mean squared error around 0.17 to 0.18. The paper also explores varying model parameters, such as the membrane time constant and maximum firing rates, and how those affect the measurement. This model is a start to modeling the responses of selective attention, however there are still improvements that can be made to better compare with the experiment, produce more accurate responses and incorporate more biologically plausible features.

\end{abstract}
\renewcommand\IEEEkeywordsname{Keywords}
\begin{IEEEkeywords}
\textit{Selective Attention, Neural Engineering Framework}
\end{IEEEkeywords}

\section{Introduction}
In the primary visual cortex, neurons in the extrastriate visual cortex have large receptive fields \cite{neural-mechanisms} \cite{cat-cortex}, which leads to computational problems when multiple stimuli fall within one receptive field. Studies have shown that in the case of two stimuli within the same receptive field, macaque monkeys are able to direct attention to one of the stimuli locations \cite{neural-mechanisms} \cite{womelsdorf}. By directing attention to different spatial locations, the attended stimulus that location can be processed selectively, while stimuli at unattended locations can be ignored. This is known as selective attention.

Selective attention is the neuronal process that allocates resources to a specific spatial location \cite{womelsdorf}. This allocation results in shifting the receptive field center toward the focus of attention and shrinking the receptive field when the attentional focus is directed into the receptive field \cite{womelsdorf}. There are two main aspects of selective attention: endogenous attention and exogenous attention. Endogenous attention is allocation of attention using a cue to a likely location for an upcoming visual target \cite{indovina}. In this case, the subject’s attention is directed to that location. Exogenous attention is when the cue presented is non-informative to the location of the upcoming visual target \cite{indovina}. In this case, the cue may be presented at the same or different location of the upcoming stimulus.

The model presented in this project aims to simulate selective attention between V1 and the middle temporal (MT) area in the visual cortex and use endogenous attention as the verification for the model. Extended and modified from a unifying mechanistic model of selective attention in spiking neurons \cite{bobier}, the model simulates a study, by Womelsdorf et al, on selective attention in macaque monkeys MT area. The study reported that when attention is directed into receptive fields of neurons in the MT area, the magnitude of the shift of the spatial-tuning functions is positively correlated with a narrowing of spatial tuning around the attentional focus \cite{womelsdorf}. The study also showed that the response is bell-shaped from the center of attention. The model in this project aim to also simulate a similar response for attend in and attend out stimulus cases.

\section{Methods}

\subsection{System Description}

\begin{figure}
	\includegraphics[width=\columnwidth]{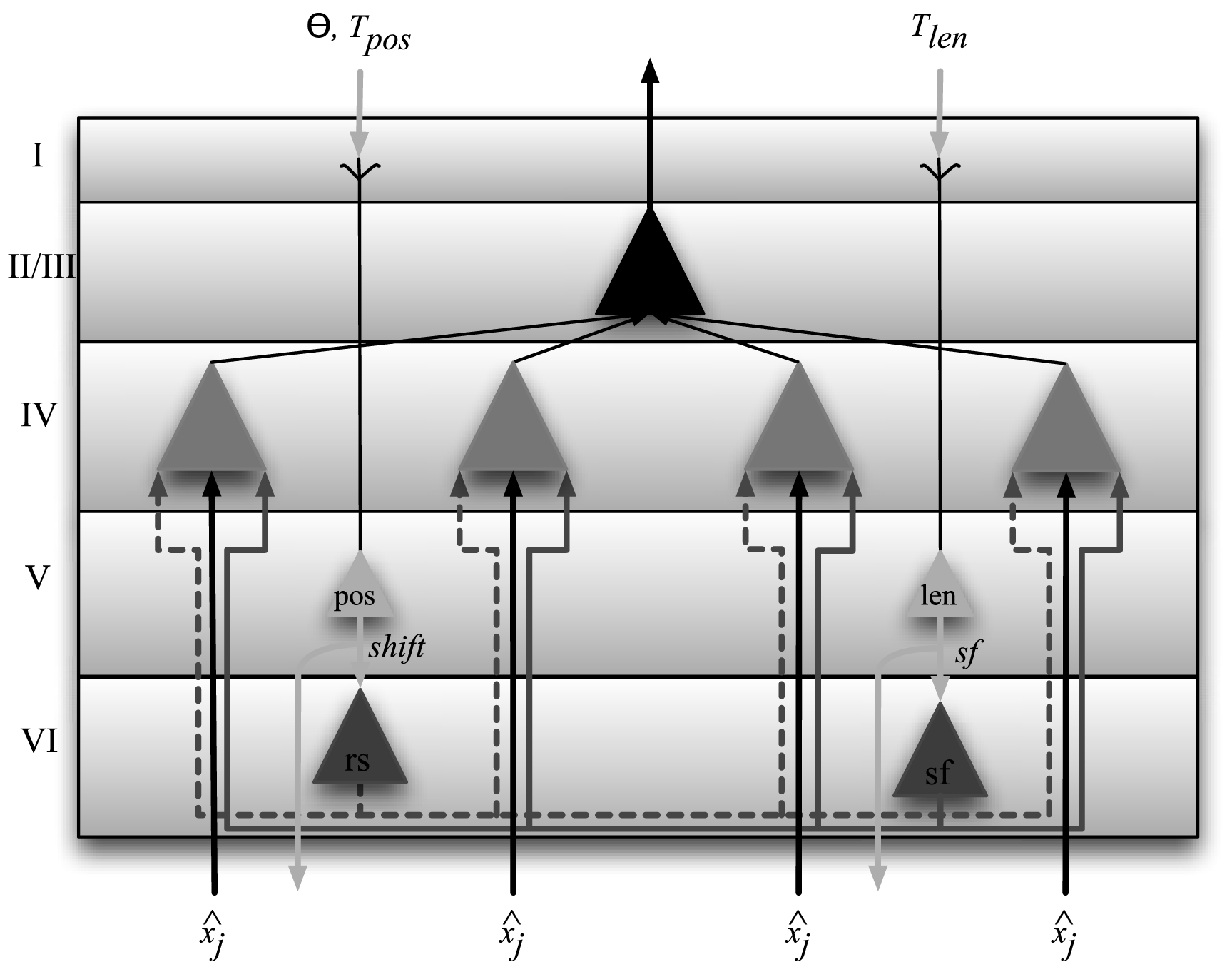}
	\caption{Attentional control through V1 layers \cite{bobier}}
	\label{fig:layers}
\end{figure}

The neural system of interest is the connection between the primary visual cortex, V1 and the middle temporal (MT) area. Visual information is passed through the magno, also called the dorsal or parietal, pathway that goes through layers VI, V, IV, III and II of V1 to MT \cite{kandel}. There are also global control signals from the pulvinar that project to the posterior inferior temporal (PIT) cortex, which are then fed into the control neurons in layers V and VI in Figure \ref{fig:layers} \cite{bobier}.

The lowest layers in V1 are layers V and VI. These layers compute local control signals, $\sigma_{att}$ and $\mu$, from the global control signals sent from PIT. $\sigma_{att}$ is the width of the local receptive field and $\mu$ is the center of the local receptive field. These local control signals guide the routing of a local portion of the attended object \cite{bobier}. After layers V and VI, layer IV is involved in selective gating of inputs. Layer IV contains nonlinear dendrites, which gate feedforward visual signals based on local control signals \cite{bobier}. Deeper in the V1 hierarchy, layer II and III process the gated visual signals. Visually responsive neurons in these layers encode the visual signal and send to MT neurons.

When extracting information that are encoded in certain layers, different nonlinear transformations are used to extract transformed versions of the signals. The transformation function for determining whether or not to gate occurs between layers VI and IV and is shown in Function 1.

\begin{lstlisting}[caption={Gating function },language=Python,escapeinside={(*}{*)}]
def gating_func(x): (*\label{lst:gating}*)
    pos = x[0] # position of V1 column
    center = x[1] # center of receptive field
    width = x[2] # radius of receptive field
    if pos > center + width or pos < center - width:
        return 0 # gating
    else:
        return 1 # no gating
\end{lstlisting}

The gating of the visual signals is determined by if the position of the visual stimulus is in the receptive field or not based on the two local control signal. Gating of the visual signals happens in layer IV and is shown in Function 2. This transformation function determines whether to encode the visual stimulus or to ignore it.

\begin{lstlisting}[caption={Signal to encode},language=Python,escapeinside={(*}{*)}]
def MT_column_func(x): (*\label{lst:signal}*)
    gating = x[0] # whether or not to gate
    stim = x[1] # encoded visual stimulus
    if gating > 0.5:
        return stim
    else:
        return 0
\end{lstlisting}

The neurons in the MT area receive input from each of the encoded visual stimulus in layer II and III of the V1 columns. However, depending on the position of the MT neurons, the responses to the visual stimulus differs. To account for this difference, a Gaussian function, shown in Equation \ref{eq:gaussian} is used, where $\mu_i$ is position of the MT neurons, $x_j$ is the position of the visual stimulus in V1 and $\sigma_{att}$ is the radius of the receptive field. The transformation is shown in Function 3.

\begin{equation}\label{eq:gaussian}
    f(\mu_i,x_j) = e^{\frac{-(\mu_i-x_j)^2}{2{\sigma_{att}}^2}}
\end{equation}

\begin{lstlisting}[caption={Response to visual stimulus},language=Python,escapeinside={(*}{*)}, label={lst:gaussian}]
def strength_func(x):
    stim = x[0] # visual stimulus
    pos = 0.5 # position of MT column
    center = x[2] # center of receptive field
    width = x[3] # radius of receptive field
    diff = (center-pos)
    f = np.exp(-(diff)**2/(2*width**2))
    return stim*f
\end{lstlisting}

\subsection{Design Specification}
In literature, the visual cortex of a cat has a maximum firing of 120 spikes per second for the most sensitive orientation \cite{firing-rate}. Additionally, the inactivation time constant, known as the refractory period is around 1-2 milliseconds in the visual cortex with total typical interspike interval between 20 to 100 ms \cite{synaptic}. The membrane time constant was found to range from 20 to 50 milliseconds for major types of central neurons \cite{time-constant}. Based on these results from literature, all neurons in the model are LIF neurons with 2 millisecond refractory periods, membrane time constants of 20 milliseconds and maximum firing rates from a uniform distribution in the range of 90 to 120 spikes per second. Since, the membrane time constants from literature vary, changing the membrane time constant are explored when evaluating the behaviour of the model.

\subsection{Implementation}
\begin{figure*}
	\includegraphics[width=\textwidth]{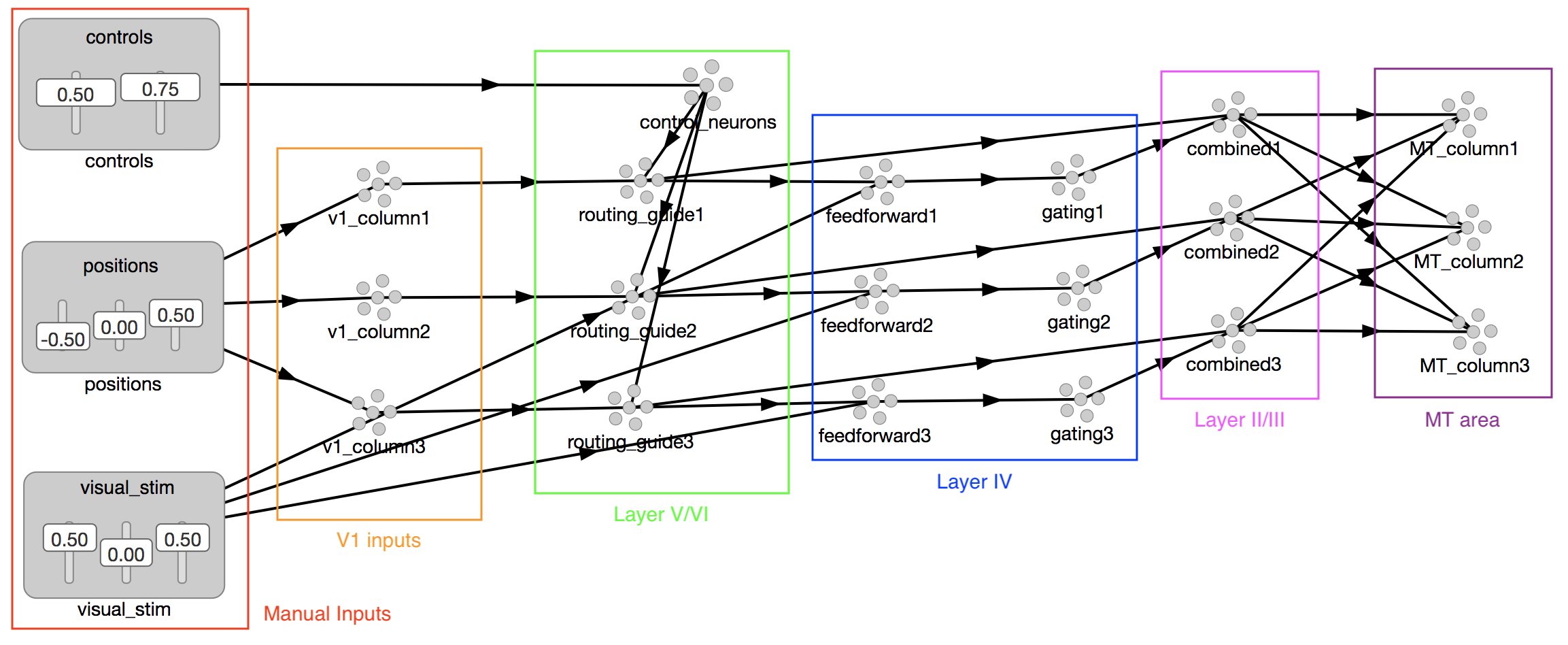}
	\caption{Implemented model in Neural Engineering Framework}
	\label{fig:model}
\end{figure*}

Given the functions of each layer and the neuron specifications, the model was implemented in the Neural Engineering Framework (NEF), shown in Figure \ref{fig:model}. The model is split into 6 different parts: the manual inputs, V1 column positions encoding, layers V and VI, layer IV, layers II and III, and MT area.

The manual inputs control the local control signals, as well as the position and strength of each visual stimulus. There are manual inputs for the local control signals because the computation of the local control signals from the global control signals are scoped out for this project. The global control signals calculate the center and radius, $\mu$ and $\sigma_{att}$, of the local receptive field depending on where the stimulus is located in the current receptive field. For this project, the center and radius are manually controlled depending on the positions of the visual stimulus. To control the positions and strengths of visual stimuli, the positions and strengths node are adjusted. Each column in the positions node corresponds to the same column in the visual stimulus node. The position and strength of a visual stimulus are both 1 dimensional.

The V1 columns are 1 dimensional and each encode the position of a visual stimulus. The connections are shown in Function \ref{lst:positions-v1}. The encoded positions are then sent to their respective groups of routing neurons in layer V and VI. Layer V and VI contain 2 dimensional control neurons that encode the local control signals and 3 dimensional routing neurons that encode the local control signals and stimulus positions. In addition to V1 columns, the control neurons also send information to the routing neurons, shown in Function \ref{lst:routing-neurons}.

\begin{lstlisting}[caption={Connection between positions node and V1 columns},language=Python,escapeinside={(*}{*)}]
nengo.Connection(positions[0], v1_column1) (*\label{lst:positions-v1}*)
nengo.Connection(positions[1], v1_column2)
nengo.Connection(positions[2], v1_column3)
\end{lstlisting}

\begin{lstlisting}[caption={Connections between V1 columns, control and routing neurons},language=Python,escapeinside={(*}{*)}]
nengo.Connection(v1_column1, routing_guide1[0]) (*\label{lst:routing-neurons}*)
nengo.Connection(v1_column2, routing_guide2[0])
nengo.Connection(v1_column3, routing_guide3[0])
nengo.Connection(control_neurons, routing_guide1[1:])
nengo.Connection(control_neurons, routing_guide2[1:])
nengo.Connection(control_neurons, routing_guide3[1:])
\end{lstlisting}

Layer IV contains 2 dimensional feedforward neurons and 1 dimensional gating neurons. The feedforward receives input from the visual stimulus and the encoded information from the routing neurons. These neurons encode the visual stimulus and whether or not the stimulus is to be gated. The connection is shown in Function \ref{lst:feedforward}. These feedforward neurons then send the encoded information to the gating neurons, shown in Function \ref{lst:gating-conn}, which determine whether to encode the visual stimulus, if not gated, or 0, if gated.

\begin{lstlisting}[caption={Connections visual stimulus, routing and feedforward neurons},language=Python,escapeinside={(*}{*)}]
nengo.Connection(visual_stim[0], feedforward1[1]) (*\label{lst:feedforward}*)
nengo.Connection(visual_stim[1], feedforward2[1])
nengo.Connection(visual_stim[2], feedforward3[1])
nengo.Connection(routing_guide1, feedforward1[0], function=gating_func)
nengo.Connection(routing_guide2, feedforward2[0], function=gating_func)
nengo.Connection(routing_guide3, feedforward3[0], function=gating_func)
\end{lstlisting}

\begin{lstlisting}[caption={Connections between feedforward and gating neurons},language=Python,escapeinside={(*}{*)}]
nengo.Connection(feedforward1, gating1, function=MT_column_func) (*\label{lst:gating-conn}*)
nengo.Connection(feedforward2, gating2, function=MT_column_func)
nengo.Connection(feedforward3, gating3, function=MT_column_func)
\end{lstlisting}

Layer II and II contain 4 dimensional combine neurons, which is an intermediary step before each visual stimulus is sent to different MT columns. These neurons take input from encoded visual stimulus in the gating neurons and position information in the routing neurons, shown in Function \ref{lst:combined}. The neurons in each 1 dimensional MT column use the encoded information from the neurons in the layer below to calculate the response to each stimulus and adds the responses to produce a total response. Function \ref{lst:columns} shows how the combine neurons use transformation functions to calculate the strengths of each input.

\begin{lstlisting}[caption={Connections between gating, routing and combined neurons},language=Python,escapeinside={(*}{*)}]
nengo.Connection(gating1, combined1[0]) (*\label{lst:combined}*)
nengo.Connection(routing_guide1, combined1[1:])
nengo.Connection(gating2, combined2[0])
nengo.Connection(routing_guide2, combined2[1:])
nengo.Connection(gating3, combined3[0])
nengo.Connection(routing_guide3, combined3[1:])
\end{lstlisting}

\begin{lstlisting}[caption={Connections between combined neurons and MT column neurons},language=Python,escapeinside={(*}{*)}]
nengo.Connection(combined1, MT_column1, function=strength_func1) (*\label{lst:columns}*)
nengo.Connection(combined2, MT_column1, function=strength_func1)
nengo.Connection(combined3, MT_column1, function=strength_func1)

nengo.Connection(combined1, MT_column2, function=strength_func2)
nengo.Connection(combined2, MT_column2, function=strength_func2)
nengo.Connection(combined3, MT_column2, function=strength_func2)

nengo.Connection(combined1, MT_column3, function=strength_func3)
nengo.Connection(combined2, MT_column3, function=strength_func3)
nengo.Connection(combined3, MT_column3, function=strength_func3)
\end{lstlisting}

\section{Results}

\subsection{Measurement}
\begin{figure}
	\includegraphics[width=\columnwidth]{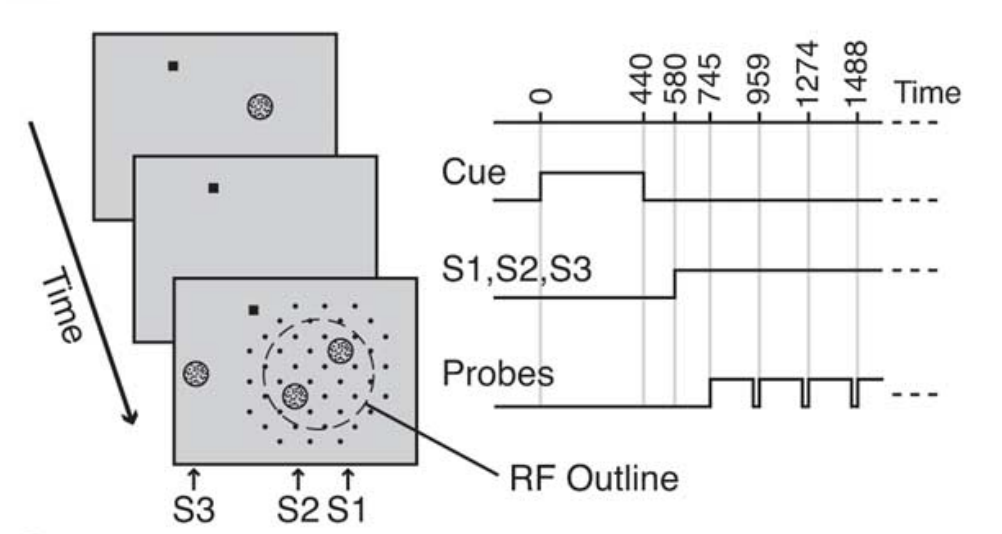}
	\caption{Layout of experiment showing succession of cue and stimuli (S1, S2 and S3) \cite{womelsdorf}}
	\label{fig:cue-stimulus}
\end{figure}

The model simulates an experiment by Womelsdorf et al. \cite{womelsdorf}. The experiment measures receptive field profiles in the macaque MT area. First, the macaque foveats on a small square, which acts as a cue, presented on a computer scene for 440 milliseconds. The cue consists of a stationary random dot pattern. After a brief blank delay, three moving random dot patterns are shown. Of the three random dot patterns, two are within the receptive field of the isolated neuron with equal eccentricity. The third random dot pattern is presented outside the receptive field in the opposite hemifield. The procedure of this experiment is shown in Figure \ref{fig:cue-stimulus}.

\begin{figure*}
	\includegraphics[width=\textwidth]{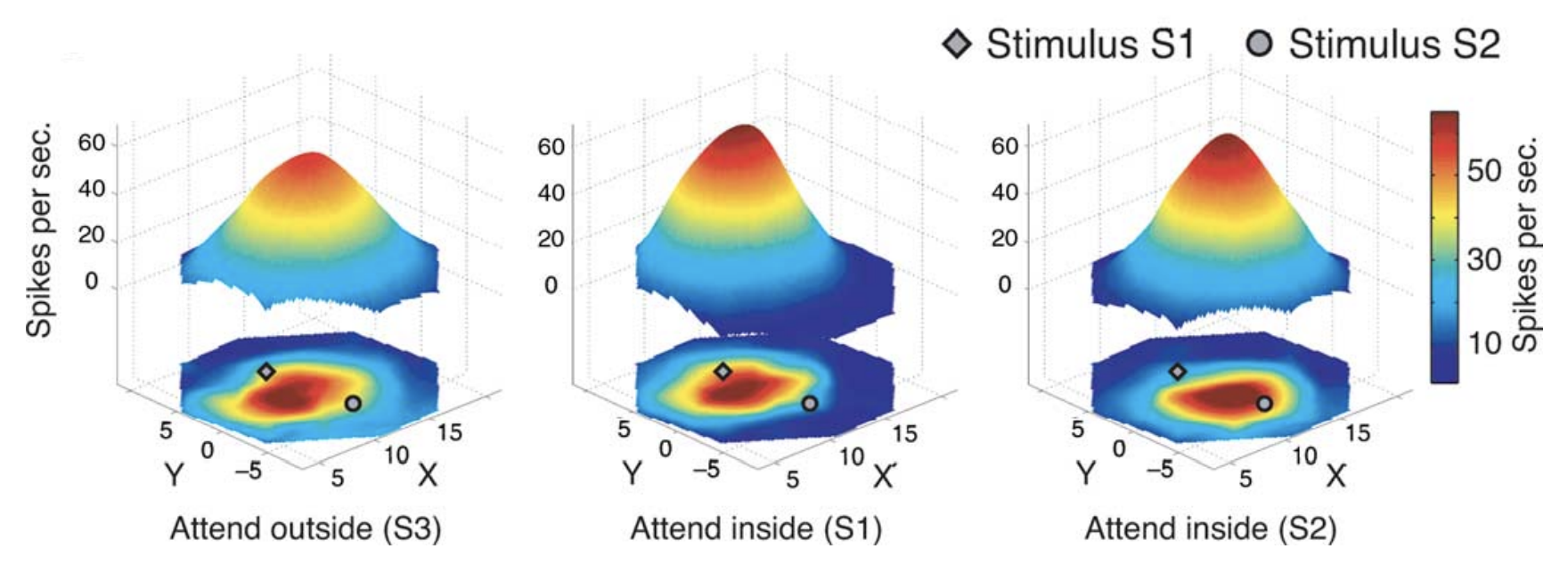}
	\caption{Receptive fields of a neuron when attention was directed outside (S3) and inside (S1, S2) of the receptive field \cite{womelsdorf}}
	\label{fig:stimulus-response}
\end{figure*}

The cue is placed where one of the three stimuli are and when the three patterns are shown, the macaque attends to the stimulus that is closest to the cue location. For the experiment, the cue is tested at all three stimuli locations and 78 probes were placed on neurons in the receptive field, shown by the black dots in Figure \ref{fig:cue-stimulus}. From the experiment, they found that the responses to the probes can be fit with a Gaussian to construct the responses in the receptive field. They also found that attending to a target stimulus inside the receptive field, S1 or S2, resulted in a shifted Gaussian peak towards the target and a smaller Gaussian width compared to when attending outside the receptive field, but no significant change in peak responses \cite{womelsdorf}. When attending to a target stimulus outside the receptive field, S3, the Gaussian peak did not shift and the width of the Gaussian stayed the same, however the peak response activity is lower than when attending inside the receptive field \cite{womelsdorf}. These results are shown in Figure \ref{fig:stimulus-response}.

For the simulation, the cue is encoded in the control neurons, with $\mu$ being the center of the Gaussian and $\sigma_{att}$ being the width of the Gaussian. When the cue is inside the receptive field, $\mu$ is set to the position of the target stimulus and $\sigma_{att}$ is set to 0.75. When the cue is outside the receptive field, $\mu$ is set to 0, which means the peak of the Gaussian stays at the center of the receptive field, and $\sigma_{att}$ is set to 1.00 since the width of the Gaussian when attending outside the receptive field. Each of the three V1 columns encodes one of the three positions, with the center being position 0. The two attend in stimuli, S1 and S2, are placed with equal eccentricity at -0.5 and 0.5. The attend out stimulus, S3, is not in this simulation since the three position columns are all within the receptive field and anything not in the receptive field will be gated. Each of the visual stimulus columns corresponds to a position column. The value of each visual stimulus represents how fast the random dot patterns are moving on average. For the simulation, the value of both the attend in stimuli are set to 0.5 to keep a small radius for the neurons to encode.

The measurement for this simulation is based on the accuracy of the representation. The experiment uses probes at 78 neurons to get the overall response in the receptive field. Contrary to the experiment, the simulation gets the response using three MT columns inside the receptive field. The placement of each MT column represents a position in the receptive field and the output of each MT column represents the response at that position. The responses of the MT columns are compared to the results in Figure \ref{fig:stimulus-response} to measure the similarity between the Gaussian responses. Additionally, since the MT column encodes the target visual stimulus, the root mean squared error between the encoded value and the target visual stimulus will be computed to measure the accuracy of the representation.

\subsection{Using starting parameters}

The initial parameters for all neurons in the simulation are LIF neurons with a refractory period of 2 milliseconds and membrane time constant of 20 milliseconds. For each ensemble, the maximum firing rate is set to a uniform distribution between 90 and 150 Hz, the number of neurons is set to 200 to 300 neurons per dimension and the radius is set to 1 for the first and last layer and 2 for the intermediate layers. The intermediate layers require a bigger radius because those layers are encoding the width of the receptive field, which can be greater than 1. The first and last layers are encoding the position and response to the visual stimulus, which are all less than 1. The simulation is ran for 1 second 10 times for S1, S2 and S3 since there is randomness in the creation of a neuron model.

\begin{figure*}
	\includegraphics[width=\textwidth]{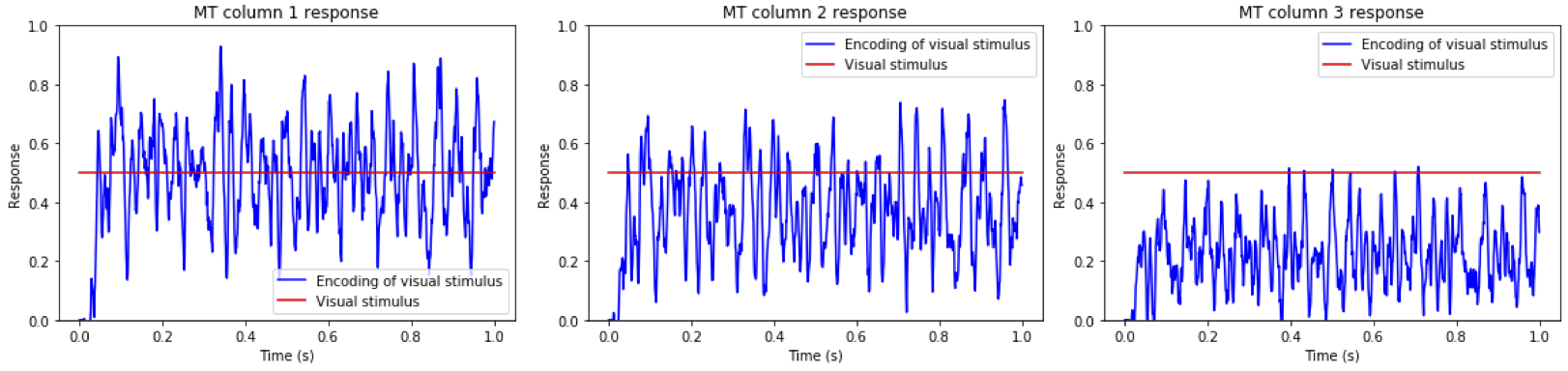}
	\caption{Response of each MT column to visual stimuli with receptive field center at -0.5}
	\label{fig:response-05}
\end{figure*}

\begin{figure*}
	\includegraphics[width=\textwidth]{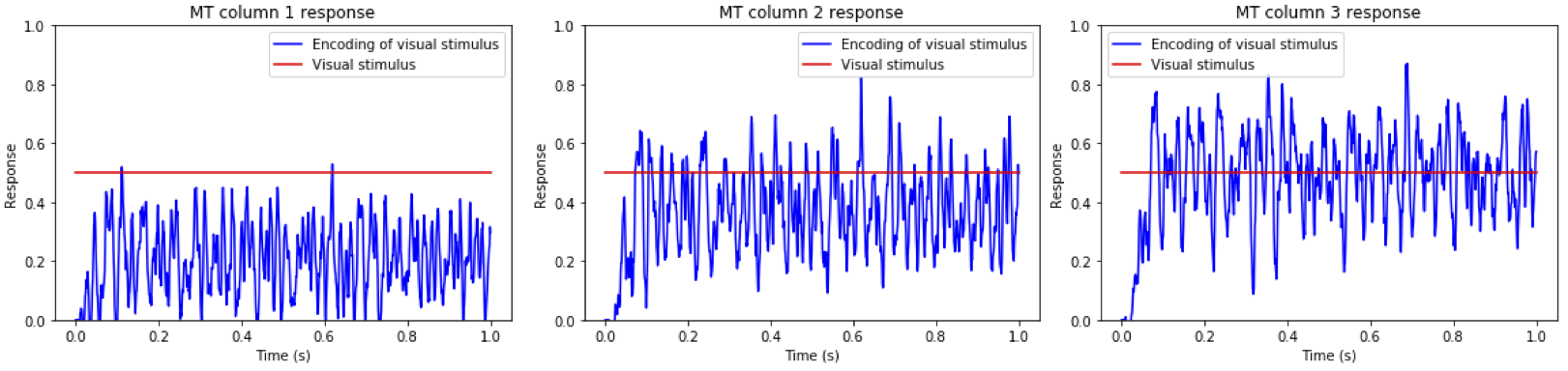}
	\caption{Response of each MT column to visual stimuli with receptive field center at 0.5}
	\label{fig:response05}
\end{figure*}

\begin{figure*}
	\includegraphics[width=\textwidth]{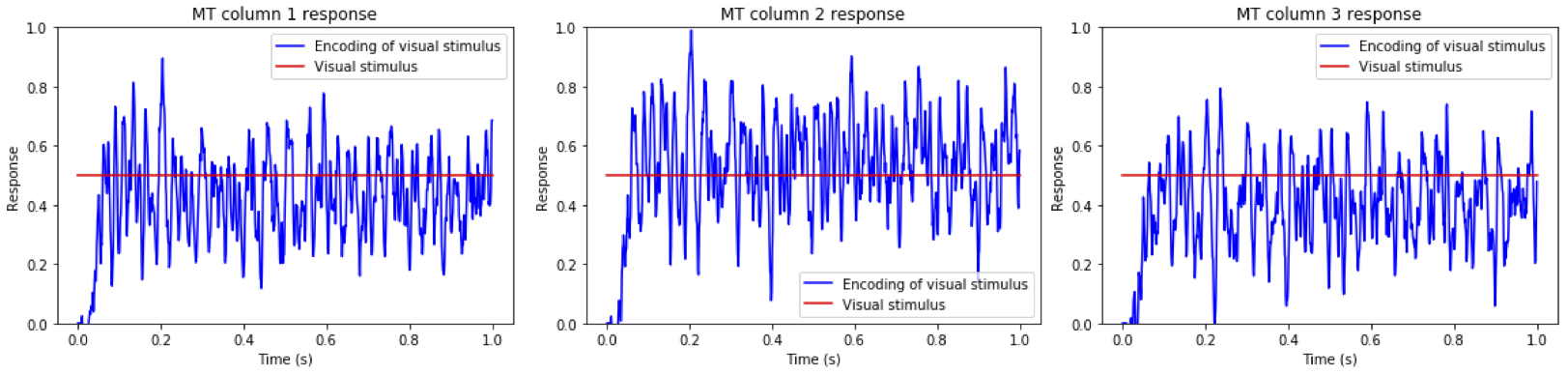}
	\caption{Response of each MT column to visual stimuli with receptive field center at 0}
	\label{fig:response0}
\end{figure*}

Using these initial parameters, the results are similar to the Gaussian when attending inside the receptive field, however, the results are slightly different than the Gaussian when attending outside the receptive field. These similarities and differences are shown in Figure \ref{fig:response-05}, \ref{fig:response0} and \ref{fig:response05}. For the attend in cases of S1 and S2, as shown in Figure \ref{fig:response-05} and \ref{fig:response05}, the shape of the responses are similar to the Gaussians in Figure \ref{fig:stimulus-response} for S1 and S2. In both the experiment and simulation, the peak of the Gaussians are shifted towards the position of where the target stimulus is located. When attending outside of the receptive field for S3, the experiment expected a lower Gaussian peak and wider width than the Gaussians for S1 and S2. Based on results from the simulation, the Gaussian peak was slightly higher than the Gaussian peaks for S1 and S2. However, the width is wider than the widths for S1 and S2 since there is a smaller difference between the responses of the adjacent MT columns. The wider width is consistent with the result from the experiment.

The shape and width of the Gaussians can be further analyzed using the standard deviation measures for the errors of the responses in each MT column. Near the peak of a Gaussian, the values are all close together, which suggests a smaller standard deviation. This is reflected in Table \ref{table:response-05} and \ref{table:response05}. For the attend in case of S1, the standard deviations in Table \ref{table:response-05} increase from MT column 1 to column 3 and similarly for the attend in case of S2, the standard deviations in Table \ref{table:response05} increase from MT column 3 to column 1. This result is consistent with Figure \ref{fig:response-05} and \ref{fig:response05} as the peaks are at MT column 1 for S1 and MT column 3 for S2. From Table \ref{table:response0}, the standard deviations for S3 are all greater than the standard deviations for S1 and S2. This suggests that the Gaussian for S3 is wider than the Gaussians for S1 and S2.

\begin{table}[h!]
\caption{RMSE values for attend in case of stimulus at -0.5 for 10 simulations}
\label{table:response-05}
\centering
\begin{tabular}{||c c c c c||}
 \hline
 Column & Average & Standard deviation & Minimum & Maximum \\ [0.5ex]
 \hline\hline
 1 & 0.184714 & 0.0167916 & 0.163839 & 0.225076 \\
 2 & 0.201632 & 0.0284993 & 0.157628 & 0.25371 \\
 3 & 0.319237 & 0.0433436 & 0.236494 & 0.386097 \\ [1ex]
 \hline
\end{tabular}
\end{table}

\begin{table}[h!]
\caption{RMSE values for attend in case of stimulus at 0.5 for 10 simulations}
\label{table:response05}
\centering
\begin{tabular}{||c c c c c||}
 \hline
 Column & Average & Standard deviation & Minimum & Maximum \\ [0.5ex]
 \hline\hline
 1 & 0.29118 & 0.0251388 & 0.240577 & 0.327063 \\
 2 & 0.183169 & 0.0207187 & 0.156774 & 0.221978 \\
 3 & 0.171846 & 0.0139234 & 0.152079 & 0.190283 \\ [1ex]
 \hline
\end{tabular}
\end{table}

\begin{table}[h!]
\caption{RMSE values for attend out case of stimulus for 10 simulations}
\label{table:response0}
\centering
\begin{tabular}{||c c c c c||}
 \hline
 Column & Average & Standard deviation & Minimum & Maximum \\ [0.5ex]
 \hline\hline
 1 & 0.230447 & 0.0377947 & 0.176116 & 0.291067 \\
 2 & 0.322718 & 0.0449662 & 0.249429 & 0.415941 \\
 3 & 0.250265 & 0.0321684 & 0.20262 & 0.330088 \\ [1ex]
 \hline
\end{tabular}
\end{table}

In addition, Table \ref{table:response-05} and \ref{table:response05} can be used to measure how accurately the MT columns are able to encode the target visual stimulus from the root mean squared error (RMSE). Similar to the pattern of standard deviations, the average RMSE values increase from MT column 1 to 3 for S1 and increase from MT column 3 to 1 for S2. The smaller the RMSE, the more accurate the representation. The smallest RMSE value for S1 and S2 corresponds to the MT column that is at the same position as the visual stimulus, which shows that the receptive field shifts towards the position of the target stimulus to get a more accurate encoding. This behaviour of the simulation is consistent with the results from the experiment.

\subsection{Adjusting parameters}
For further evaluation, the following parameters are varied to analyze the behaviour of the simulation model: membrane time constant, maximum firing rate, number of neurons per dimension and radius. All parameters, except for the varied parameter, are kept as the initial parameters so that all changes are associated with the varied parameter. For each varied parameter, the model ran for 1 second 10 times with $\mu=0.5$ and another 10 times with $\mu=-0.5$ due to randomness in the creation of a neuron model.

\subsubsection{Membrane time constant}

\begin{figure*}[h!]
	\includegraphics[width=\textwidth]{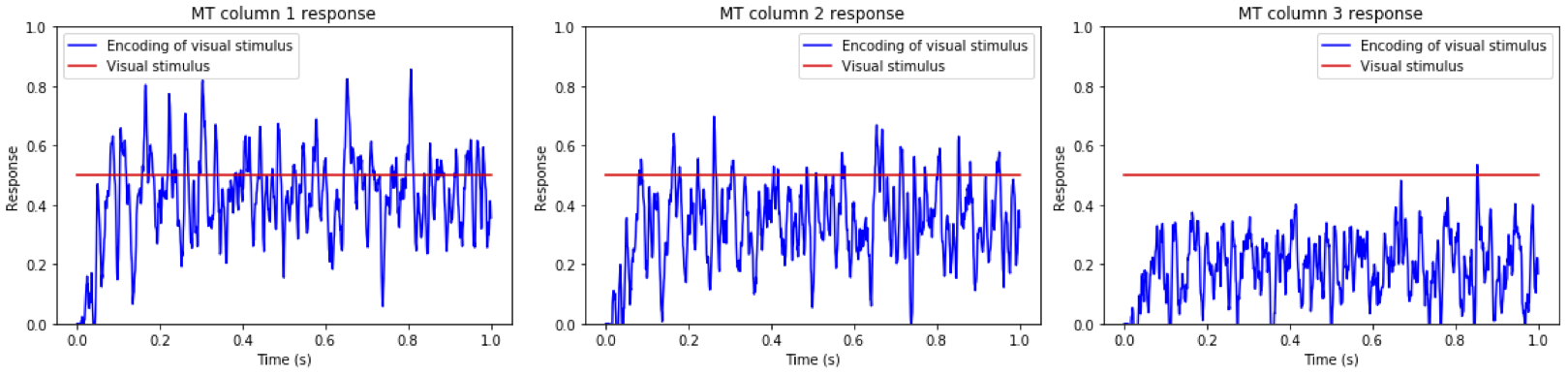}
	\caption{Response of each MT column to visual stimuli with receptive field center at -0.5 with $\tau_{RC}=50ms$}
	\label{fig:response-05-rc}
\end{figure*}

\begin{figure*}[h!]
	\includegraphics[width=\textwidth]{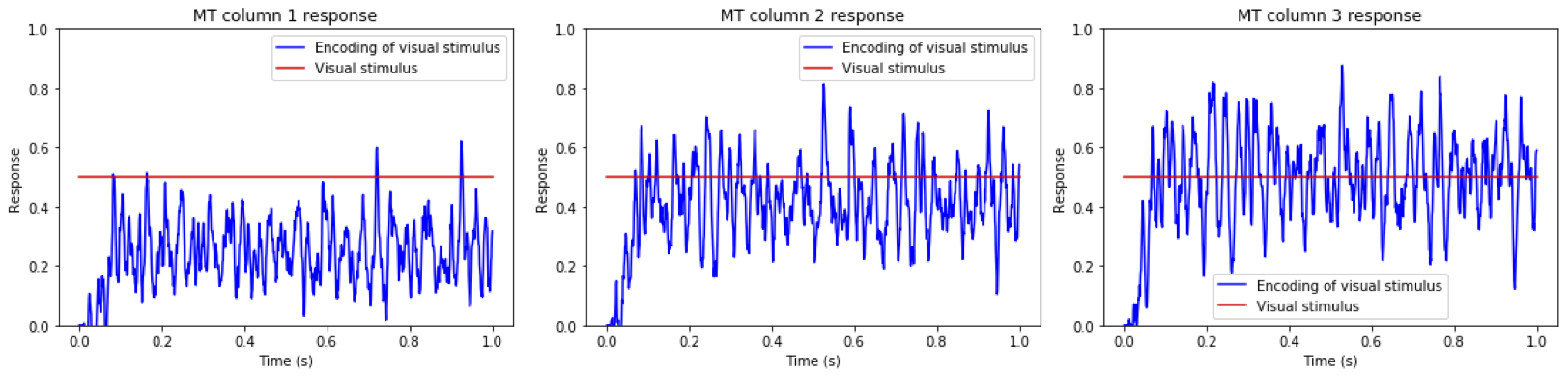}
	\caption{Response of each MT column to visual stimuli with receptive field center at 0.5 with $\tau_{RC}=50ms$}
	\label{fig:response05-rc}
\end{figure*}

Based on literature, membrane time constants vary from 20 to 50 milliseconds. Since 20 milliseconds is used as an initial parameter, to further evaluate the behaviour of the simulation, the membrane time constant, $\tau_{RC}$, is changed to 50 milliseconds. The MT column responses are shown in Figure \ref{fig:response-05-rc} and \ref{fig:response05-rc} and the RMSE results for 10 simulations are in Table \ref{table:response-05-rc} and \ref{table:response05-rc}.

From Figure \ref{fig:response-05-rc} and \ref{fig:response05-rc}, the responses for each MT column is very similar to the results with a membrane time constant of 20 milliseconds. The similarity can be verified using Table \ref{table:response-05-rc} and \ref{table:response05-rc}. The average RMSE for each MT column is only slightly better, approximately 0.016, than the average RMSE when membrane time constant is 20 milliseconds. This suggests that the membrane time constant has little effect on the behaviour of the simulation. As well, variations from 20 to 50 milliseconds in membrane time constants between neurons are biologically plausible and does not affect the performance of selective attention. The model is consistent with this fact and shows that varying the membrane time constant to 50 milliseconds does not affect the response of the MT columns.

\begin{table}[h!]
\caption{RMSE values for attend in case of stimulus at -0.5 for 10 simulations with $\tau_{RC}=50ms$}
\label{table:response-05-rc}
\centering
\begin{tabular}{||c c c c c||}
 \hline
 Column & Average & Standard deviation & Minimum & Maximum \\ [0.5ex]
 \hline\hline
 1 & 0.169989 & 0.0117122 & 0.15294 & 0.186731 \\
 2 & 0.201751 & 0.0309556 & 0.160126 & 0.257653 \\
 3 & 0.314501 & 0.0525482 & 0.222169 & 0.391263 \\ [1ex]
 \hline
\end{tabular}
\end{table}

\begin{table}[h!]
\caption{RMSE values for attend in case of stimulus at 0.5 for 10 simulations with $\tau_{RC}=50ms$}
\label{table:response05-rc}
\centering
\begin{tabular}{||c c c c c||}
 \hline
 Column & Average & Standard deviation & Minimum & Maximum \\ [0.5ex]
 \hline\hline
 1 & 0.299613 & 0.0308468 & 0.248119 & 0.358184 \\
 2 & 0.196585 & 0.0374033 & 0.161649 & 0.277934 \\
 3 & 0.169712 & 0.0229544 & 0.142258 & 0.217545 \\ [1ex]
 \hline
\end{tabular}
\end{table}

\subsubsection{Maximum firing rate}
\begin{figure*}[h!]
	\includegraphics[width=\textwidth]{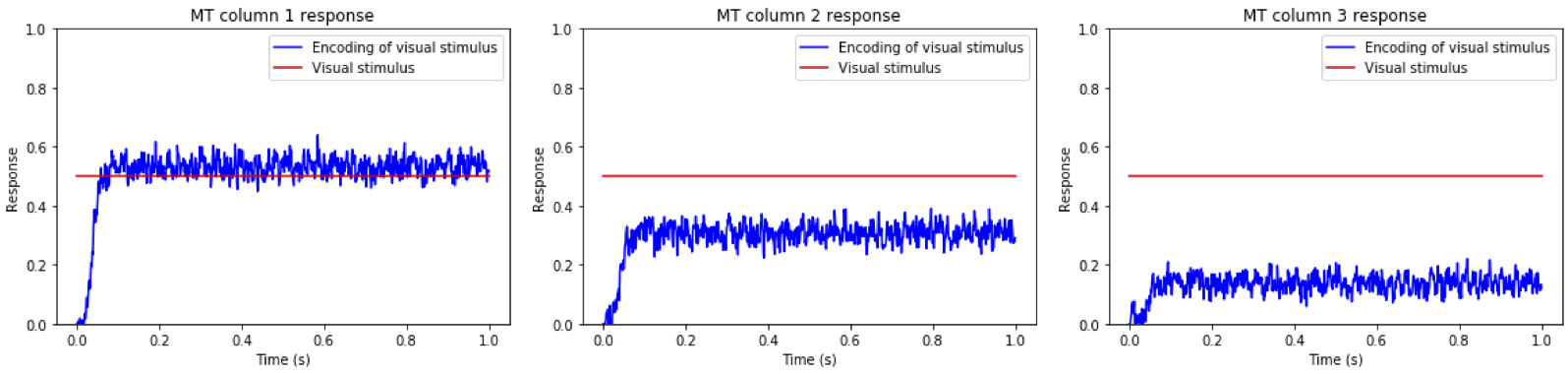}
	\caption{Response of each MT column to visual stimuli with receptive field center at -0.5 with $max_{firing rate} = Uniform(200,400)$}
	\label{fig:response-05-fr}
\end{figure*}

\begin{figure*}[h!]
	\includegraphics[width=\textwidth]{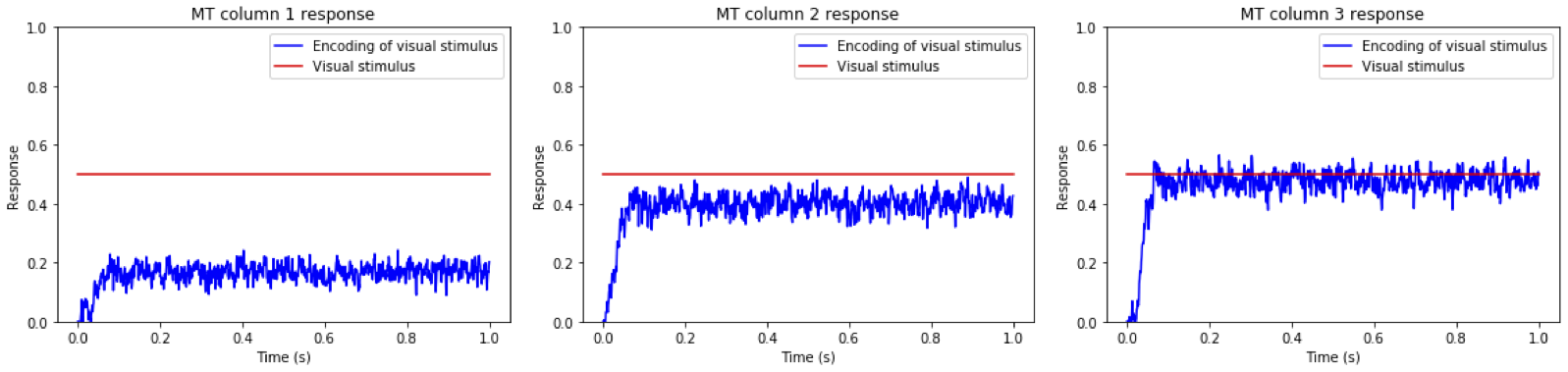}
	\caption{Response of each MT column to visual stimuli with receptive field center at 0.5 with $max_{firing rate} = Uniform(200,400)$}
	\label{fig:response05-fr}
\end{figure*}

Using maximum firing rates from 90 to 150 are biologically plausible for these neurons, however, the results in are very noisy, as indicated by the high RMSE values. One way to lessen the noise and lower the RMSE values is to increase the maximum firing rates. The maximum firing rates were adjusted to a uniform distribution between 200 to 400 Hz, which are the default maximum firing rates in NEF. The MT column responses are shown in Figure \ref{fig:response-05-fr} and \ref{fig:response05-fr} and the RMSE results for 10 simulations are in Table \ref{table:response-05-fr} and \ref{table:response05-fr}.

Both Figure \ref{fig:response-05-fr} and \ref{fig:response05-fr} show a less noisy signal compared to the results using maximum firing rates between 90 and 150 Hz. The average RMSE for the MT column at the same position as the target stimulus is significantly lower than the average RMSE with the initial maximum firing rates by approximately 0.06. There is also a greater difference in RMSE between the two MT columns closer to the target stimulus and a smaller difference in RMSE between the middle MT column and furthest MT column, which is closer to the features of a Gaussian. Gaussians have a steeper slope near the peak and more tapered slope further from the peak. This shape is more prominent with higher maximum firing rates, however is not biologically plausible based on literature. Based on this result, the model still needs improvements to get more accurate Gaussian shaped responses when attending within the receptive field.

\begin{table}[h!]
\caption{RMSE values for attend in case of stimulus at -0.5 for 10 simulations with $max_{firing rate} = Uniform(200,400)$}
\label{table:response-05-fr}
\centering
\begin{tabular}{||c c c c c||}
 \hline
 Column & Average & Standard deviation & Minimum & Maximum \\ [0.5ex]
 \hline\hline
 1 & 0.12041 & 0.0200961 & 0.0926795 & 0.159034 \\
 2 & 0.150369 & 0.0495381 & 0.0902525 & 0.280328 \\
 3 & 0.303619 & 0.073508 & 0.188586 & 0.40876 \\ [1ex]
 \hline
\end{tabular}
\end{table}

\begin{table}[h!]
\caption{RMSE values for attend in case of stimulus at 0.5 for 10 simulations with $max_{firing rate} = Uniform(200,400)$}
\label{table:response05-fr}
\centering
\begin{tabular}{||c c c c c||}
 \hline
 Column & Average & Standard deviation & Minimum & Maximum \\ [0.5ex]
 \hline\hline
 1 & 0.316525 & 0.0426524 & 0.244202 & 0.373887 \\
 2 & 0.171923 & 0.0616212 & 0.102591 & 0.312941 \\
 3 & 0.11796 & 0.0172861 & 0.0962067 & 0.161336 \\ [1ex]
 \hline
\end{tabular}
\end{table}

\subsubsection{Number of neurons per dimension}
The average number of neurons in V1 for a galago is 35 million and the average number of neurons in MT is 1.6 million \cite{collins}. For the model, the total number of neurons used for the simulation is 8200 for the V1 layers and 900 neurons for the MT area. Since there are 35 million and 1.6 million neurons in V1 and MT respectively, it is biologically plausible that the V1 and MT layers in model can have more neurons to encode information. Thus, the number of neurons per dimension were adjusted to 500. This resulted in a total of 31 000 neurons for the model, with 16 000 neurons in the V1 layers and 1500 neurons in the MT area. The RMSE results for 10 simulations are in Table \ref{table:response-05-nn} and \ref{table:response05-nn}. Based on the RMSE results, there is no difference when increasing the number of neurons.

\begin{table}[h!]
\caption{RMSE values for attend in case of stimulus at -0.5 for 10 simulations with $num_{neurons}=500$ per dimension}
\label{table:response-05-nn}
\centering
\begin{tabular}{||c c c c c||}
 \hline
 Column & Average & Standard deviation & Minimum & Maximum \\ [0.5ex]
 \hline\hline
 1 & 0.168458 & 0.0130997 & 0.151311 & 0.200504 \\
 2 & 0.190773 & 0.0263891 & 0.159197 & 0.253793 \\
 3 & 0.296992 & 0.0473238 & 0.247796 & 0.419308 \\ [1ex]
 \hline
\end{tabular}
\end{table}

\begin{table}[h!]
\caption{RMSE values for attend in case of stimulus at 0.5 for 10 simulations with $num_{neurons}=500$ per dimension}
\label{table:response05-nn}
\centering
\begin{tabular}{||c c c c c||}
 \hline
 Column & Average & Standard deviation & Minimum & Maximum \\ [0.5ex]
 \hline\hline
 1 & 0.315767 & 0.0527126 & 0.20165 & 0.388242 \\
 2 & 0.195296 & 0.040469 & 0.154633 & 0.283632 \\
 3 & 0.177385 & 0.0174921 & 0.152879 & 0.213773 \\ [1ex]
 \hline
\end{tabular}
\end{table}

\subsubsection{Radius}
Except for the first and last layer of the model, the other neurons have radiuses of 2 to encode values greater than 1. For instance, the width of the receptive field, $\sigma_{att}$, can be greater than 1. The first and last layer does not exceed 1 because the visual stimulus and the visual stimulus positions are all within 1. However, as the radius increases, the RMSE also increases linearly. This is because as the radius increases, there is more area where the neurons are not tuned to. Since the simulations above do not have $\sigma_{att}$ greater than 1, the radius can be reduced to 1 to avoid tuning values greater than 1. The RMSE results for 10 simulations are in Table \ref{table:response-05-r} and \ref{table:response05-r}.

Similar to increasing the maximum firing rates, decreasing the radius also improves the RMSE for the MT column that is closest to the target stimulus. However, unlike the result from increasing the maximum firing rate, the RMSE for the second closest MT column to the target stimulus is decreased to be slightly larger than the closest MT column, minimizing the difference between the two columns. This is not desired as it is different from the shape of a Gaussian, which only has a steep slope near the peak. Decreasing the radius also decreases the standard deviations for all the columns, which suggests that more neurons are now tuned to values within a radius of 1.

\begin{table}[h!]
\caption{RMSE values for attend in case of stimulus at -0.5 for 10 simulations with $radius=1$}
\label{table:response-05-r}
\centering
\begin{tabular}{||c c c c c||}
 \hline
 Column & Average & Standard deviation & Minimum & Maximum \\ [0.5ex]
 \hline\hline
 1 & 0.133066 & 0.0195901 & 0.110138 & 0.181869 \\
 2 & 0.136355 & 0.0145683 & 0.112102 & 0.154539 \\
 3 & 0.309206 & 0.039069 & 0.245352 & 0.363603 \\ [1ex]
 \hline
\end{tabular}
\end{table}

\begin{table}[h!]
\caption{RMSE values for attend in case of stimulus at 0.5 for 10 simulations with $radius=1$}
\label{table:response05-r}
\centering
\begin{tabular}{||c c c c c||}
 \hline
 Column & Average & Standard deviation & Minimum & Maximum \\ [0.5ex]
 \hline\hline
 1 & 0.332018 & 0.040232 & 0.284684 & 0.391706 \\
 2 & 0.157829 & 0.0353274 & 0.117858 & 0.219992 \\
 3 & 0.143153 & 0.035082 & 0.108155 & 0.220612 \\ [1ex]
 \hline
\end{tabular}
\end{table}

\section{Discussion}
The model simulates a selective attention experiment by Womelsdorf et al. \cite{womelsdorf}. The responses from the three MT columns are used to compare with the experiment results of the Gaussian receptive field model and how accurately each MT column encodes the target visual stimulus. Based on the results, there are three improvements that can be made to the model to better compare with the experiment, produce more accurate responses and incorporate more biologically plausible features. There are always more improvements that can be made on this model, however, these three improvements do not involve additional subsystems to the model.

The first improvement is adding more MT columns to better analyze the pattern of responses. For the experiment, 78 probes are used to measure the response of the Gaussian receptive field in the MT area. More MT columns can be implemented in either 1 dimension or 2 dimensions like the experiment. Adding more MT columns allow the results to be compared more easily and gives visualization to the slope of the responses.

The second improvement is to reduce the amount of large oscillations in the output of the MT columns. As shown in Figure \ref{fig:response-05}, \ref{fig:response05} and \ref{fig:response0}, there are many large oscillations. When the maximum firing rates are increased to a uniform distribution between 200 and 400, there are less large oscillations, as shown in Figure \ref{fig:response-05-fr} and \ref{fig:response05-fr}. However, the maximum firing rates cannot be increased because increasing the maximum firing rates are not biologically plausible for V1 neurons. The question for this improvement is: how to adjust the model to produce similar results without changing the firing rates? Further research would have to go into this improvement to get more accurate responses.

Lastly, the final improvement is to more accurately represent the width of the attended receptive field, $\sigma_{att}$, by calculating the shrinkage based on the amount that was shifted for attend in cases, Equation \ref{eq:width-in}, or by calculating the spread for attend out cases, Equation \ref{eq:width-out} \cite{womelsdorf}. Currently, $\sigma_{att}$ is set to 0.75 if the target is within the receptive field and 1.0 if the target is outside the receptive field. This is not an accurate measure of $\sigma_{att}$, however having $\sigma_{att}$ smaller when attending within the receptive field than when attending outside gives an approximate response of the output since this is what happens in biology.

\begin{equation} \label{eq:width-in}
    shrinkage = \sqrt{1-shift}
\end{equation}

\begin{equation} \label{eq:width-out}
    \sigma_{Att} = \sigma_{R}\sqrt{\frac{1}{shift}-1}
\end{equation}

\section{Conclusions}
From the results using initial parameters, the simulation is able to produce Gaussian shaped responses similar to the experiment. When varying the parameters of membrane time constant and number of neurons per dimension, there were little to no change in the results. Changing the maximum firing rate to a uniform distribution between 200 and 400 Hz produced the most significant change that matches the results from the experiment the most accurately. However, this change is not biologically plausible for V1 neurons. Lastly, the radius was decreased to improve RMSE. This decreased the RMSE, however, the result was not desirable since the two columns closest to the position of the target visual stimulus had little difference, which is different from the shape of a Gaussian. Overall, this model is a start to representing the selective attention responses in MT columns. More improvement is needed for the model to improve both the response accuracy and for the model to be more biologically plausible.

\bibliographystyle{ieee}

\end{document}